\documentclass[referee]{aa}     
\usepackage{graphicx}
\usepackage{txfonts}

\begin{document}

\title{Time delay of SBS 0909+532}

\author{A. Ull\'an\inst{1} \and L. J. Goicoechea\inst{1} \and A. P. Zheleznyak\inst{2}
	\and E. Koptelova\inst{3} \and V. V. Bruevich\inst{3} \and T. Akhunov\inst{4}
	\and O. Burkhonov\inst{4}}

\offprints{A. Ull\'an}

\institute{Departamento de F\'{\i}sica Moderna, Universidad de Cantabria, 
	Avda. de Los Castros s/n, 39005 Santander, Spain\\
	\email{aurora.ullan@postgrado.unican.es, goicol@unican.es} 
	\and
	Institute of Astronomy of Kharkov National University, 
	Sumskaya 35, 61022 Kharkov, Ukraine\\
	\email{zheleznyak@astron.kharkov.ua} 
	\and 
	Sternberg Astronomical Institute, Universitetski pr. 13, 
	119992 Moscow, Russia\\
	\email{koptelova@xray.sai.msu.ru, bruevich@sai.msu.ru} 
	\and
	Ulug Beg Astronomical Institute of Uzbek Academy of 
	Science, Astronomicheskaya. Str. 33, 700052 Tashkent, Republic 
	of Uzbekistan\\
	\email{talat77@rambler.ru, boa@astrin.uzsci.net}}
		  
\date{Accepted January 12, 2006}

\titlerunning{Time delay of SBS 0909+532}
\authorrunning{A. Ull\'an et al.}

\abstract{
The time delays between the components of a lensed quasar are basic tools to analyze the 
expansion of the Universe and the structure of the main lens galaxy halo. In this paper, we 
focus on the variability and time delay of the double system SBS 0909+532A,B as well as the
time behaviour of the field stars. We use $VR$ optical observations of SBS 0909+532A,B and the 
field stars in 2003. The frames were taken at Calar Alto, Maidanak and Wise observatories, and 
the $VR$ light curves of the field stars and quasar components are derived from aperture and 
point--spread function fitting methods. We measure the $R$--band time delay of the system from 
the $\chi^2$ and dispersion techniques and 1000 synthetic light curves based on the observed 
records. One nearby field star (SBS 0909+532c) is found to be variable, and the other two nearby 
field stars are non--variable sources. With respect to the quasar components, the $R$--band 
records seem more reliable and are more densely populated than the $V$--band ones. The observed 
$R$--band fluctuations permit a pre--conditioned measurement of the time delay. From the $\chi^2$ 
minimization, if we assume that the quasar emission is observed first in B and afterwards in A 
(in agreement with basic observations of the system and the corresponding predictions), we obtain 
$\Delta \tau_{BA}$ = $-$ 45 $^{+ 1}_{-11}$ days (95\% confidence interval). The dispersion technique 
leads to a similar delay range. A by--product of the analysis is the determination of a totally 
corrected flux ratio in the $R$ band (corrected by the time delay and the contamination due to the 
galaxy light). Our 95\% measurement $\Delta m_{BA}$ = $m_B(t + \Delta \tau_{BA}) - m_A(t)$ = 
0.575 $\pm$ 0.014 mag is in excellent agreement with previous results from contaminated fluxes at 
the same time of observation.
}

\maketitle

\section{Introduction}

The system SBS 0909+532 was discovered by Stepanyan et al. (1991). Some years later,
a collaboration between the Hamburger Sternwarte and the Harvard--Smithsonian Center
for Astrophysics resolved the system into a pair of quasars (A and B) with a direct 
$R$--band flux ratio (at the same time of observation) $\Delta m$ = $m_B - m_A$ = 0.58 
mag and a separation of about 1\farcs1 (Kochanek et al. 1997). The direct $R$--band
flux ratio was not consistent with the direct flux ratios at other wavelengths: $\Delta m$ 
= 0.31 mag in the $I$ band and  $\Delta m$  = 1.29 mag in the $B$ band. From observations
with the 4.2 m William Herschel Telescope, a Spanish collaboration got spectra for each 
component of the system. The data showed that the system consists of two quasars with
the same redshift  ($z_s$ = 1.377) and identical spectral distribution, supporting the 
gravitational lens interpretation of SBS 0909+532 (Oscoz et al. 1997). Oscoz et al. (1997) 
detected a \ion{Mg}{ii} doublet in absorption at the same redshift ($z_{abs}$ = 0.83) in 
both components, and they suggested that the absorption features were associated with the 
photometrically unidentified lensing galaxy. Through a singular isothermal sphere (SIS) 
lens model, the authors also inferred the first constraint on the time delay between the 
components: $|\Delta \tau_{BA}| \leq$ 140 days, where $\Delta \tau_{BA}$ is the delay of B 
with respect to A and the Hubble constant is assumed to be $H_0$ = 70 km s$^{-1}$ Mpc$^{-1}$. 

In recent years, Lubin et al. (2000) indicated the possible nature of the main deflector 
(early--type galaxy) and confirmed its redshift ($z_d$ = 0.830). Leh\'ar et al. (2000) reported 
on a program including Hubble Space Telescope (HST) observations of SBS 0909+532. They 
discovered the main lens galaxy between the components, which has a large effective radius, 
with a correspondingly low surface brightness. This lens galaxy is closer to the brightest 
component (A), which is not in contradiction with SIS--like lens models when the farther and 
fainter component (B) is stronger affected by dust extinction (see below). The colors of 
the lens are consistent with those of an early--type galaxy at redshift 0.83. Assuming a 
singular isothermal ellipsoid (SIE) model, Leh\'ar et al. predicted a time delay $\Delta 
\tau_{BA}$ in the range [$-$ 10, $-$ 87] days ($H_0$ = 70 km s$^{-1}$ Mpc$^{-1}$). At a given 
emission time, the sign "$-$" means that the corresponding signal is observed first in B and 
later in A. The COSMOGRAIL collaboration provided the distribution of predicted time delays of 
the system (Saha et al. 2005). In their histogram (Fig. 10 of Saha et al.), there are two 
features: the main feature is an asymmetric peak around $-$ 80 days and the secondary one is 
another asymmetric peak around $-$ 45 days. Therefore, if the COSMOGRAIL predictions are right, 
the time delay is very probably of 2--3 months (component B leading component A), but we 
cannot rule out a delay of about one and a half months. On the other hand, the flux ratio 
anomaly pointed by Kochanek et al. (1997) was confirmed and accurately studied by Motta et al. 
(2002) and Mediavilla et al. (2005), who reported the existence of differential extinction in 
the main lens galaxy. Chartas (2000) and Page et al. (2004) also studied the system in the 
X--ray domain. 

Time delays are basic tools to discuss the present expansion rate of the Universe and the 
structure of the main lens galaxy haloes (e.g., Refsdal 1964; Kochanek, Schneider \& Wambsganss 
2004), so that variability studies are crucial. While some time delays have been measured from 
radio light curves (PKS 1830$-$211: Lovell et al. 1998; Q0957+561: Haarsma et al. 1999; 
B0218+357: Biggs et al. 1999; B1600+434: Koopmans et al. 2000; B1422+231: Patnaik \& Narasimha 
2001; B1608+656: Fassnacht et al. 2002) or X--ray variability (e.g., Q2237+0305: Dai et al. 
2003), an important set of delays are based on optical monitoring of gravitationally lensed 
quasars. Optical frames taken at Apache Point Observatory, Fred Lawrence Whipple Observatory 
and Teide Observatory were used to estimate a 14--month delay for the double system Q0957+561 
(e.g., Pelt et al. 1996; Kundi\'c et al. 1997; Serra--Ricart et al. 1999; Ovaldsen et al. 2003). 
Although the time delay of this first multiple quasar has been confirmed through independent 
observations, the measurement is only 5\% accurate, or equivalently, there is an uncertainty of 
about 20 days (Goicoechea 2002). The Tel--Aviv University (TAU) group have recently determined 
the time delay between the two components of HE 1104$-$1805 (Ofek \& Maoz 2003). The TAU delay 
of HE 1104$-$1805 disagrees with the earlier estimation by Gil--Merino, Wisotzki \& Wambsganss 
(2002), but it is in excellent agreement with the determination by Wyrzykowski et al. (2003). 
Schechter et al. (1997) measured two delays for the quadruply imaged quasar PG 1115+080. The 
Belgian--Nordic collaboration carried out a very intense activity during the past five years. 
They participated in several monitoring projects and measured several time delays at optical 
wavelengths: B1600+434 (Burud et al. 2000), HE 2149$-$2745 (Burud et al. 2002a), RXJ 0911.4+0551 
(Hjorth et al. 2002), SBS 1520+530 (Burud et al. 2002b) and FBQ 0951+2635 (Jakobsson et al. 
2005). The formal accuracies of these 5 estimations range from 5 to 25\% (the 1$\sigma$ error 
bars vary from 4 to 24 days). Kochanek et al. (2005) also measured the time delays between the 
components of the quadruple quasar HE 0435$-$1223.

The aim of this paper is to present $VR$ observations of SBS 0909+532 in 2003 conducted by 
the University of Cantabria (UC, Spain), the Institute of Astronomy of Kharkov National 
University (IAKhNU, Ukraine), the Sternberg Astronomical Institute (SAI, Russia) and the 
Ulug Beg Astronomical Institute of Uzbek Academy of Science (UBAI, Uzbekistan). We also
present TAU observations of the field stars in 2003, which have been kindly made available to 
us. This new optical monitoring campaign was carried out at the Calar Alto Observatory (Spain), 
the Maidanak Observatory (Uzbekistan) and the Wise Observatory (Israel), and the frames were 
taken with the 1.5 m Spanish telescope, the 1.5 m AZT$-$22 telescope at Mt. Maidanak and the
Wise Observatory 1 m telescope (Section 2). In Section 3, we describe the methodology to 
obtain the fluxes of the quasar components and the field stars. The $VR$ light curves are also 
shown in Section 3. Section 4 is devoted to the time delay estimation from the light curves 
of A and B (quasar components) in the $R$ band. Finally, in Section 5 we summarize our 
conclusions and discuss the feasibility of an accurate determination of the cosmic expansion 
rate and the surface density in the main lensing galaxy.

\section{Observations}

\begin{figure}
\centering
\includegraphics[angle=-90,width=10cm]{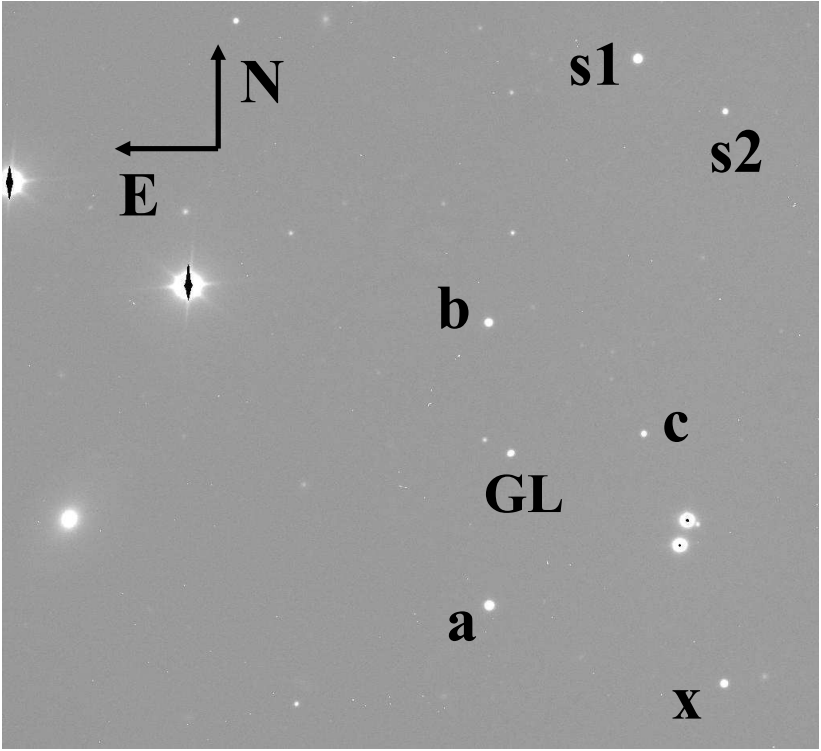}
\caption{Calar Alto image of SBS 0909+532, showing a FOV of $\sim 7\arcmin \times 7\arcmin$.
The field contains the gravitationally lensed quasar ("GL") and six bright and non--saturated  
stars ("a--c", "s1--s2" and "x"). The nearby ("a--c") and relatively far ("s1--s2") field stars 
were introduced by Kochanek et al. (1997) and Nakos et al. (2003), respectively. A sixth field 
star ("x") is also included in the FOV.}
\end{figure}

We have three different sets of frames for SBS 0909+532. The first set of optical frames cover 
the period between 2003 March 4 and June 2, and they are part of a UC project to test the 
feasibility of quasar monitoring programs through 1$-$2 m telescopes in Spain (Ull\'an 2005). 
These observations were made with the 1.52 m Spanish telescope at Calar Alto Observatory (EOCA), 
Almeria, Spain (see Ziad et al. 2005 for a site--testing on Calar Alto). The EOCA is equipped 
with a Tektronics 1024$\times$1024 CCD detector, which has pixels with a physical size of 
24 $\mu$m, giving a 0.4 arcsec pixel$^{-1}$ angular scale. The gain is 6.55 e$^{-}$/ADU and the 
readout noise is 6.384 e$^{-}$. During this first monitoring, exposures in the $V$ and $R$ 
Johnson--Cousins filters were taken every night when clear, what makes a total of 20 observing 
nights. Bad weather in 2003 March and April prevented us from achieving a very dense sampling. 
For each monitoring night we have three consecutive frames on each filter, i.e., three 300 s 
exposures in the $V$ passband and three 180 s exposures in the $R$ passband. Those were the 
maximum exposure times to avoid saturation of selected stars in the field. In Figure 1 we show 
a typical frame. In this typical exposure, half a dozen bright and non--saturated stars were 
fitted within the field of view (FOV). Following the notation of Kochanek et al. (1997), the 
FOV included the gravitationally lensed quasar ("GL") and nearby field stars "a" (South), "b" 
(North) and "c" (West). The FOV also included two stars that were introduced by Nakos et al. 
(2003) and were labelled as "s1" and "s2". These two stars are placed relatively far from the 
gravitational lens system, and they appear close to the North--West edge of the frame (see Fig. 
1). A sixth star ("x") appears close to the South--West edge of the typical frame.

\begin{figure}
\centering
\includegraphics[angle=0,width=12cm]{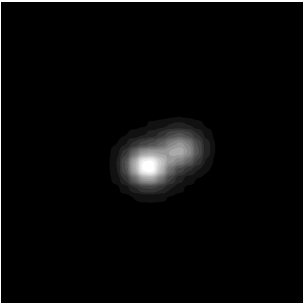}
\caption{Maidanak zoomed--in image of SBS 0909+532 (about 8\arcsec on a side). We choose one 
of the frames taken with the best seeing conditions, and then we expand the portion occupied by 
the two quasar components A (the brightest object) and B (the faintest object), i.e., the GL in 
Fig. 1. To avoid the pixellation effect when a subframe is expanded, the brightness distribution 
is also smoothed.}
\end{figure}

The second set of observations include frames in February 2003 as well as during April--May 
and October--November 2003. The total number of nights is 18. In this second program the 
images were taken with the 1.5 m AZT$-$22 telescope at Maidanak Observatory (Uzbekistan), 
with near diffraction--limited optics and careful thermo--stabilization, which allow for 
high--angular--resolution imaging. The AZT$-$22 telescope has a LN--cooled (liquid nitrogen 
cooled) CCD--camera, SITe- 005 CCD, manufactured in Copenhagen (Denmark). For this camera, the 
imaging area is split into 2000$\times$800 pixels, where the pixel size is 15 $\mu$m and the
intrinsic angular scale is 0.26 arcsec pixel$^{-1}$. The frames were taken in the $R$ 
Bessel filter, which corresponds approximately to the $R$ Johnson--Cousins passband. The poor 
tracking system of this telescope allows only exposures up to 3 minutes. To obtain sufficiently 
high photometric accuracy, we took several frames each observation night. With respect to the 
rectangular FOV of the telescope, the North/South coverage was 2.5 times smaller than the 
East/West one, so the "s1", "s2" and "x" stars were not included within the FOV. Figure 2 shows 
a zoomed--in image made from one of the best frames in terms of seeing. There are two close
quasar components, but the very faint galaxy is not apparent. The observations at Mt. Maidanak 
are part of IAKhNU, SAI and UBAI projects to follow up the variability of gravitationally 
lensed quasars.

\begin{table}
\centering
\begin{tabular}{ccc}
\hline\noalign{\smallskip}
Observatory (Telescope) & Frames/night (Filter) & Observation Periods \\
\noalign{\smallskip}\hline\noalign{\smallskip}
Calar Alto (1.5 m) & 3 $\times$ 300 s ($V$) + 3 $\times$ 180 s ($R$) & March--June \\
Maidanak (1.5 m) & (3--11) $\times$ 180 s ($R$)  & February, April--May, October--November \\
Wise (1.0 m) & 1 $\times$ 420 s ($R$) & 22 unevenly distributed nights \\
\noalign{\smallskip}\hline
\end{tabular}
\caption{Observations of SBS 0909+532 in 2003.\label{tbl-1}}
\end{table}

For the past six years the TAU group have been monitoring several gravitationally lensed quasars
with the Wise Observatory 1 m telescope. The targets are mainly monitored in the Johnson--Cousins
$R$--band, and the frames are obtained with a cryogenically cooled Tektronix 
1024$\times$1024--pixel back--illuminated CCD. The angular scale is 0.7 arcsec per pixel. This
pixel scale and the median seeing ($FWHM$) of about 2\arcsec\ do not allow resolving most of the
lensed objects, e.g., SBS 0909+532. However, the frames of SBS 0909+532 in 2003 are characterized
by wide FOVs, which incorporate the "a--c", "s1--s2" and "x" stars. This fact permits to do 
differential photometry between several pairs of field stars, and thus, to test the reliability 
of the Calar Alto and Maidanak records. 

The pre--processing of the images included the usual bias subtraction, flat fielding using sky 
flats, sky subtraction and cosmic ray removal by using the Image Reduction and Analysis 
Facility (IRAF) and Munich Image Data Analysis System (MIDAS) environments. Some details about the
whole observational campaign are included in Table 1 (observatories, telescopes, frames/night,
filters and observation periods).

\section{Photometry and $VR$ light curves}
 
Due to the small angular separation between the two lensed components, about 1\farcs1 (Kochanek 
et al. 1997), the photometry of SBS 0909+532 is a difficult task. This task is also complicated 
by the presence of the main lensing galaxy between the components, which could make the 
computation of individual fluxes even harder. In general, aperture photometry does not work, so 
we must look for better approaches. An initial issue is to decide about the inclusion or not 
inclusion of a photometric model for the lensing galaxy. In principle, when computing the fluxes 
of SBS 0909+532 we may use a galaxy model derived from the HST images of the system. The galaxy 
model could also be inferred from the best images in terms of seeing. Once the relevant 
information on the galaxy is known, we would apply a PSF fitting method to all optical images, 
setting the galaxy properties to those derived from the HST or the best--quality images, and 
allowing the remaining parameters to vary (e.g., McLeod et al. 1998; Ull\'an et al. 2003). 
Magain, Courbin \& Sohy (1998) also presented an alternative task (deconvolution) that combines 
all the frames obtained at different epochs to determine the numerical light distribution of the 
lensing galaxy as well as the positions of the point--like sources (quasar components), since 
these parameters do not vary with time. The flux of the point--like sources are allowed to vary 
from image to image, which produces the light curves. However, these and other procedures have a 
reasonable limitation: they only work well when the galaxy light has a significant contribution 
to the crowded regions in the individual frames. For a very faint galaxy in a standard (i.e., not 
superb) frame, there is confusion between galaxy signal and noise, so the use of a given galaxy 
model could lead to biased fluxes of the components. The biases will depend on the quality of 
the image (seeing, signal--to--noise ratio, etc), which must produce artificial variability 
superposed to the real one. On the other hand, the use of a direct PSF fitting method 
(neglecting the galaxy brightness) leads to contaminated fluxes of the components. But if the 
galaxy is very faint, the contaminations will be small. Moreover, the variation of the quasar 
fluxes, seeing conditions, etc, will cause fluctuations in the contaminations, which are 
expected to be below the typical contamination levels. For standard frames of a quasar lensed 
by a very faint extended object, it is really difficult to choose between both approaches (with 
and without galaxy). 

Most of the Calar Alto and Maidanak individual frames of SBS 0909+532 do not show evidences 
for a galaxy brightness profile. This fact is due to the faintness of the galaxy, as we corroborate here
below. If we consider an hypothetical astronomer that neglects the galaxy brightness and 
does direct PSF fitting (without taking into account the galaxy when doing the computation of the 
fluxes), it is possible to attain a rough estimation of the maximum contamination from the galaxy 
to the closest component A (at 0\farcs4 from the centre of the deflector). We take into account the 
paper about 10 lens systems by Leh\'ar et al. (2000), where, in Table 3, we can find the best 
available photometric and astrometric (HST) data of SBS 0909+532. The authors were able to trace the 
galaxy light in the $H$ passband, by measuring its position and brightness. If we use the colors in 
the same table, we conclude that $m_{gal} \sim$ 19 mag and $m_A \sim$ 16 mag in the $I$ band 
(near--IR), and $m_{gal} >$ 20.4 mag and $m_A \sim$ 16.7 mag in the $V$ optical band. Therefore, as 
the $R$ filter is placed just between the $I$ filter and the $V$ one, we may assume that $m_{gal} - 
m_A \sim$ 3.5 mag in the $R$ band. The difference of 3.5 mag is consistent with a ratio of fluxes 
$F_{gal}/F_A$ of about 1/25. Thus, in the case of QSO 0957+561 we found a $R$--band ratio of fluxes 
$F_{gal}/F_A$ of about 1/2.5 (Ull\'an et al. 2003), and now we have $F_{gal}/F_A \sim$ 1/25, what 
explains our unsuccessful efforts when measuring the flux of the lens galaxy in standard frames. As 
a result of that, in an extreme case (when direct PSF fitting leads to a magnitude $m_{A + gal}$ 
instead of $m_A$, i.e., all the galaxy light is included in the profile of the A component) we find 
a relationship: $m_A = m_{A + gal} + F_{gal}/F_A$, where the true flux (in magnitudes) $m_A$ differs 
from the contaminated flux through direct PSF fitting ($m_{A + gal}$) in a quantity $F_{gal}/F_A$. 
This maximum contamination of A would be only of 40 mmag, and the real contamination of both 
components will be less than our upper limit. The artificial fluctuations (caused by variable 
contamination) will be even smaller than the typical contamination levels, so we expect they will 
not play an important role in analyses of quasar variability (e.g., time delay estimates). 
 
In order to derive the light curves of the components A and B, we decide to use a direct PSF fitting 
method and do not consider the galaxy brightness in the fits. The key idea of this procedure is to 
obtain the different fluxes we are interested in by using a PSF that comes from a bright star in the 
field common to all frames. The point--like objects (quasar components and stars) are modelled by 
means of the empirical PSF. Hence, we do not use a theoretical PSF (i.e., Gaussian distribution, 
Lorentzian distribution, etc), but the two--dimensional profile of a star in this field (a PSF star). 
Apart from a PSF star, we also need a reference star to do differential photometry and to obtain 
relative fluxes $m_A - m_{ref}$ and $m_B - m_{ref}$. The good behaviour of the reference star is 
usually checked by using a control star, so the fluxes $m_{con} - m_{ref}$ are expected to agree with 
a constant level. Nevertheless, since the $R$--band flux ratio is discussed in Section 5, we also want 
to obtain a rough estimation of the contaminations from this direct technique. With this aim, a 
deconvolution technique (Koptelova et al. 2005) is also applied to a set of frames with good seeing 
and signal. The selected frames are fitted to a model including the galaxy, and thus, we are able to 
obtain a few clean fluxes of components A and B and compare them with the corresponding contaminated 
fluxes (through a direct PSF fitting). The averaged contaminations are used in Section 5.

\begin{figure}
\centering
\includegraphics[angle=0,width=10cm]{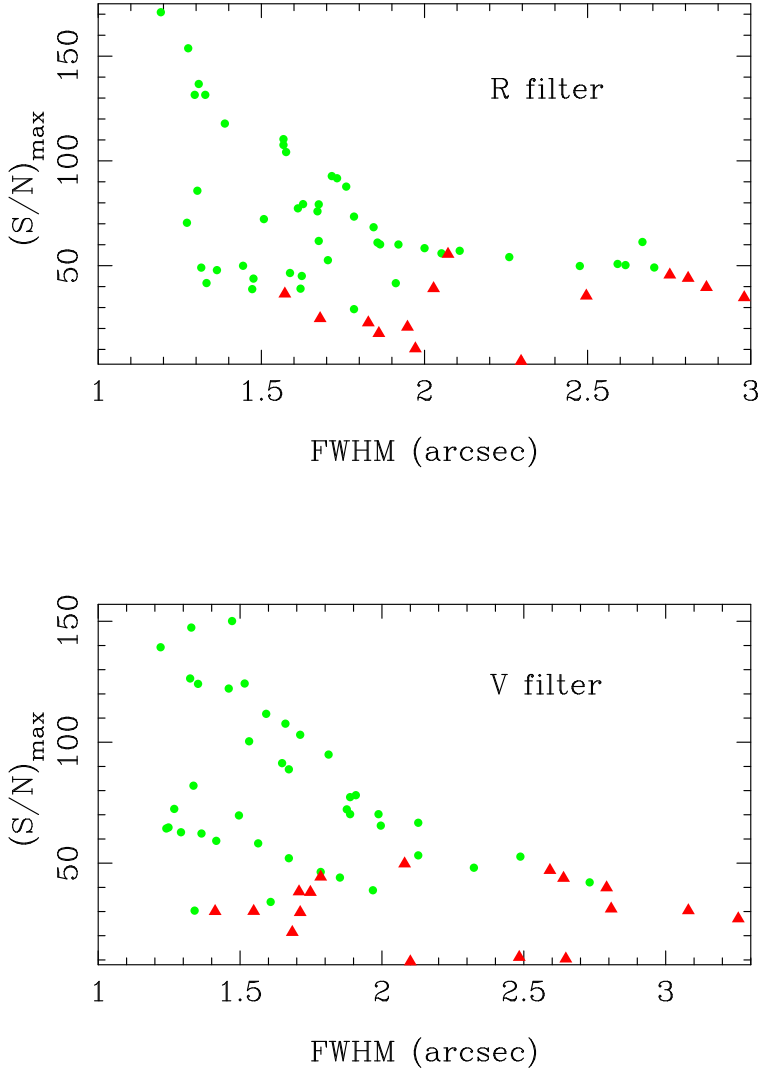}
\caption{$(S/N)_{max}$ versus $FWHM$ for the Calar Alto frames in the $R$ filter (top panel) and 
the $V$ filter (bottom panel). The circles and triangles represent the frames with good and bad 
post--fit residues, respectively (see main text).}
\end{figure}

\begin{figure}
\centering
\includegraphics[angle=-90,width=8cm]{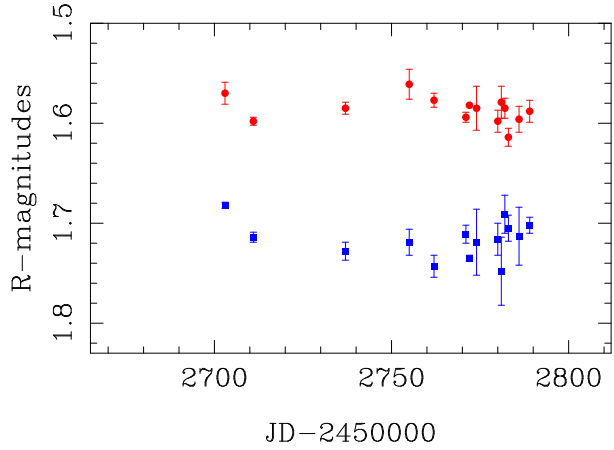}
\caption{Calar Alto light curves of the double quasar in the $R$ filter. We use the "b" star as the 
reference object, because it is confirmed as a non--variable star in Fig. 7. The circles are the 
fluxes $y_A$ ($y_A = m_A - m_b$) and the squares are the fluxes $y_B -$ 0.45 mag ($y_B = m_B - m_b$). 
The A light curve has a final decline, while the B light curve ends with a rise.}
\end{figure}

\begin{figure}
\centering
\includegraphics[angle=-90,width=8cm]{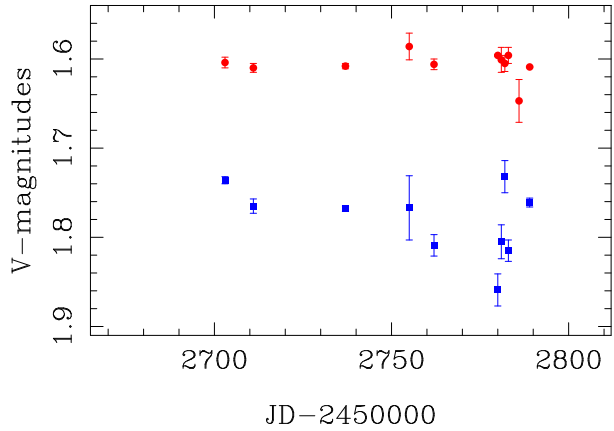}
\caption{Calar Alto light curves of the double quasar in the $V$ filter. We use the non--variable 
star "b" as reference object (see Fig. 7). The circles are the fluxes $y_A$ and the squares are the 
fluxes $y_B -$ 0.65 mag.}
\end{figure}

\begin{figure}
\centering
\includegraphics[angle=-90,width=8cm]{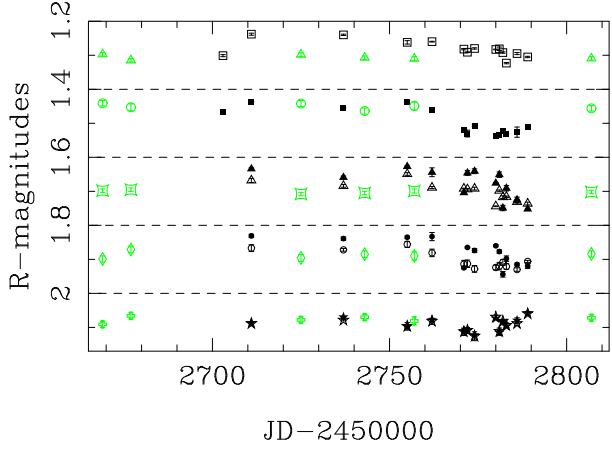}
\caption{Calar Alto light curves of the stars in the $R$ filter. The Calar Alto fluxes are checked 
from six Wise frames. Filled and open symbols are associated with PSF fitting and aperture, respectively. 
First, the non--variable star "b" (see Fig. 7) is taken as the reference object. The top box incorporates 
the behaviour of $y_a$ + 2.15 mag ($y_a = m_a - m_b$): Calar Alto (open squares) and Wise (open triangles). 
The second box (under the top box) contains the fluxes $y_c$ ($y_c = m_c - m_b$): Calar Alto (filled 
squares) and Wise (open circles). The third and fourth boxes include the $y_{s1}$ + 2.38 mag ($y_{s1} = 
m_{s1} - m_b$) and $y_{s2}$ + 0.29 mag ($y_{s2} = m_{s2} - m_b$) records, respectively. In the third box: 
Calar Alto (filled and open triangles) and Wise (open astroids), whereas in the fourth box: Calar Alto 
(filled and open circles) and Wise (open rhombuses). Second, we plot the $m_{s2} - m_{s1} -$ 0.20 mag 
records in the bottom box: Calar Alto (filled and open star symbols) and Wise (open crosses).}
\end{figure}

\subsection{Calar Alto frames and light curves}

We adopt a model of the system including two point--like sources and a constant background. The 
model is fitted to each image by adjusting its 7 free parameters (two--dimensional positions of A and 
B, instrumental fluxes of both components and background) to minimize the sum of the square residuals, 
as described in McLeod et al. (1998) and Leh\'ar et al. (2000). We use windows of 64$\times$64 
pixels. Each empirical PSF is a subframe of 64$\times$64 pixels around the PSF star (the "a" star in 
Fig. 1), while the lens system is analyzed from a subframe of the same size, but centered on the 
double quasar. The instrumental fluxes of the "b", "c", "s1" and "s2" stars are also inferred from 
64$\times$64 pixels windows centered on them. We initially focus on the nearby field stars, and take the 
"b--c" stars as the control--reference objects. The "a" object is the brightest star in the "a--c" 
triangle, and "b" and "c" were spectroscopically identified by Kochanek et al. (1997) and Zickgraf et 
al. (2003): "b" is a FG star, whose spectrum includes the G--band and \ion{Ca}{ii} H--K lines, and "c" 
is a M3 star. The $R - I$ and $B - R$ colors of the brightest component (A) and the "a" star are similar, 
the colors of the faintest component (B) are close to the colors of the "b" star, and the "c" star has 
colors different to those of the components and the "a--b" stars (see Table 1 of Kochanek et al. 1997). 
On the other hand, after checking the PSFs of the three nearby field stars ("a--c"), we do not find 
significant differences between them. This suggests that the global shape of the PSFs around the lens 
system does not depend on the position and color of the point--like objects, so the PSF of the "a" star 
seems to be a reliable tracer of the PSF associated with any point--like object in the region of 
interest. 

As a first attempt for obtaining light curves we use the "b" and "c" stars as the control and reference 
objects, respectively. Unfortunately, we find clear evidences in favour of variability of the "c" star, 
since the three curves $m_A - m_c$, $m_B - m_c$ and $m_b - m_c$ have a similar global behaviour. This 
fact forces us to rule out the "c" star as a reference--control object and, thus, to take the "a" and "b"
nearby field stars as the control and reference point--like sources, respectively. In the next subsection, 
we analyze the Maidanak--Wise fluxes $m_a - m_b$ and show that both stars ("a" and "b") are non--variable 
objects. This result permits to assure the good behaviour of "b". The Calar Alto fluxes $m_a - m_b$ 
are not included in the analysis, since most Calar Alto data disagree with the Maidanak--Wise 
common level of flux. We found an anomaly in the behaviour of the Calar Alto relative fluxes for widely 
separate stars (see below), so only the relative fluxes for neighbouring point--like objects are reliable
photometric measurements. Fortunately, the comparison between the quasar components and the "b" nearby 
reference star seems to be a feasible approach.

After applying the photometric method to the three individual frames for each filter and night (see 
Table 1), we obtain three different measurements of $y_A = m_A - m_b$ and $y_B = m_B - m_b$ in the $V$ and 
$R$ passbands for each night. To test the reliability of the instrumental fluxes of A and B, we analyse 
the residues in each residual frame. A residual frame is an image after subtracting the fitted background 
and point--like objects (PSF fitting method). More properly, we focus on the residual subframe occupied by 
the system, and then we estimate the residue--to--signal ratio ($R/S$) in each pixel of interest. A $R/S$ 
value less than 10\% is acceptable, so a subframe with at least 90\% of pixels having acceptable residues 
is considered to be related to reliable photometric solutions. Thus, we classify the individual fits in 
two categories: fits leading to $<$ 90\% of pixels having acceptable residues (bad fits, unreliable 
results) and good fits that are associated with reliable results ($\geq$ 90\% of pixels having acceptable 
residues). As a complementary test, we study the relation between the quality of the fits (in terms of 
post--fit residues) and two relevant parameters (image quality). The signal--to--noise at the brightest 
pixel of the lens system, $(S/N)_{max}$, and the seeing, $FWHM$ (in \arcsec), are the two parameters to 
compare with the fit quality. Some kind of correlation between good fits and good images is expected. In 
Figure 3 we draw the $(S/N)_{max}$--$FWHM$ plots for frames in the $R$ filter (top panel) and the $V$ 
filter (bottom panel). Circles and triangles represent good and bad fits, respectively. The plots in 
Fig. 3 indicate that the good fits correspond to images with high or moderate $(S/N)_{max}$ ($\geq$ 30). 
Moreover, at moderate $(S/N)_{max}$ ($\sim$ 30--50), most of the good fits seem to be associated with a 
relatively good seeing ($<$ 2 \arcsec). To obtain a robust photometry, we finally discard the frames 
corresponding to the triangles in Fig. 3. For each filter and night, if there are two or three good 
frames (good fits), then we get mean values of $y_A$ and $y_B$, and compute standard deviation of means 
as errors. We only consider relative fluxes with uncertainties $\leq$ 40 mmag. 

Now we plot $y_A$ (circles) and $y_B - 0.45$ mag (squares) in Figure 4 ($R$--band fluxes). If we 
concentrate our attention in the period with the best sampling (after day 2755), the A light curve shows 
a moderate decline and the B record shows a moderate rise. Indeed it seems that the "b" star is a good 
reference object (constant flux), since there is no zero--lag global correlation between $y_A$ and $y_B$. 
In Figure 5 we show the light curves $y_A$ and $y_B -$ 0.65 mag in the $V$ passband. In this case we have 
a total of 11 points for the A component (circles) and 10 points for the B component (squares). The 
$V$--band and $R$--band light curves of the A component are consistent with each other. A final moderate 
decline appears in both curves. The situation is more confused for the B component. The $R$--band final 
rise is not clearly reproduced in the $V$ band, and the $V$--band final measurements could have 
underestimated formal errors. We note the relative faintness of B in the $V$ band ($\Delta m \sim$ 0.8 
mag), and thus, the possibility of systematic uncertainties when the PSF fitting method is applied at 
some epochs. The data in both optical filters are available at http://grupos.unican.es/glendama/.

After presenting the records of the double quasar, we concentrate on the Calar Alto light curves of 
the field stars that were previously introduced by Kochanek et al. (1997) and Nakos et al. (2003), i.e., 
$y_a = m_a - m_b$, $y_c = m_c - m_b$, $y_{s1} = m_{s1} - m_b$ and $y_{s2} = m_{s2} - m_b$. There are no
previous studies on the variability of the nearby field stars "a--c". On the other hand, the farther
field stars ("s1--s2") were verified to be non--variable by using 76 Wise frames taken from 1999 
December 24 to 2002 March 3 (Nakos et al. 2003). As Nakos et al. (2003) found that "s1" and "s2" seem 
to be useful reference stars, we check the behaviour of "s1--s2" in 2003. The PSF of the stars in the 
surroundings of the double quasar could slightly differ from the PSF of the "s1--s2" stars in a 
relatively far region. Therefore, we must be careful when obtaining the instrumental fluxes of the 
farther stars. To detect possible anomalies caused by a mismatch between the brightness profile of the 
"a" star and the PSF of "s1--s2", the light curves $y_{s1}$ and $y_{s2}$ are derived from both PSF 
fitting and aperture methods. The records $y_a$, $y_c$, $y_{s1}$ and $y_{s2}$ in the $R$ filter are 
depicted in Figure 6. To guide the eyes, we use some offsets and dashed horizontal lines and put all 
the relative records of each pair within a box. Filled and open symbols are associated with PSF fitting 
and aperture, respectively. The top box includes the $y_a$ + 2.15 mag fluxes (open squares). The second, 
third and fourth boxes (under the top one) correspond to the $y_c$ (filled squares), $y_{s1}$ + 2.38 
mag (filled and open triangles) and $y_{s2}$ + 0.29 mag (filled and open circles) records, respectively. 
As most of the stars are brighter than the quasar components (A and B) and they are far from other 
objects, the typical formal errors in the stellar fluxes are clearly less than the typical uncertainties 
in the fluxes of the components (these are usually fainter and are placed in a crowded region). The 
stellar error bars in Fig. 6 are often smaller than the sizes of the associated symbols.

When doing aperture photometry on six $R$--band Wise frames covering the first semester of 2003, we 
obtain a $y_a$ + 2.15 mag light curve (open triangles in the top box of Fig. 6) that disagrees with the 
Calar Alto trend in the overlap period (between days 2710 and 2760). In the next subsection, we show 
that the Wise and Maidanak brightnesses are constant and consistent with each other, so the Calar Alto 
values of $y_a$ are not true fluxes, but anomalous results. On the contrary, the Wise light curve $y_c$ 
(open circles in the second box of Fig. 6) agrees with the Calar Alto curve in the overlap period. From 
the Wise frames we confirm the flux level during the high--state of "c". Unfortunately, the 
small--amplitude variability of "c" (rms fluctuation of $\sim$ 8 mmag) cannot be confirmed from the Wise 
data. The rms fluctuation of the Wise fluxes ($\sim$ 9 mmag) is very similar to the Calar Alto variation, 
but the formal errors are relatively large ($\sim$ 10 mmag). Moreover, there are no Wise frames in 2003 
May (around the day 2780) and, thus, we cannot check (via Wise data) the reliability of the Calar Alto 
dip in $y_c$ (80--100 mmag). However, the flux of the "c" star at day 2793 in the $V$ band confirms the 
existence of a transition from the low--state to the high--state, which is finished at days 2800--2810 
(see the last open circle in the second box of Fig. 6). For the "s1--s2" stars, which are as far from 
star "b" as star "a" is, we again find a disagreement between the Calar Alto trends and the Wise records 
(open astroids and rhombuses in the third and fourth boxes of Fig. 6). Although aperture curves are 
closer to the Wise behaviours, we cannot fairly reproduce the Wise data. Some probes with the "x" 
star (using $y_x = m_x - m_b$) also indicate that the Calar Alto and Wise behaviours disagree. It seems 
that the differential photometry between widely separate stars may lead to meaningless results, and only 
the relative fluxes for neighbouring objects are reliable. To test this conclusion, apart from the 
successful results through the neighbouring stars "b" and "c", we also analyze the differential 
photometry between the pair "s1--s2" (see Fig. 1). The curves  $m_{s2} - m_{s1} -$ 0.20 mag are depicted 
in the bottom box of Fig. 6: Calar Alto (filled and open star symbols) and Wise (open crosses). In the 
overlap period (from day 2710 to day 2760), there is a reasonable agreement between the results from both 
observatories, and the Calar Alto measurements seem to be quite reliable. From the Calar Alto frames, 
both photometric techniques are consistent with each other, but a constant flux cannot explain the 
observations. When we fit the data sets to a constant, our best solutions are characterized by $\chi^2 
\sim$ 162 (PSF fitting) and $\chi^2 \sim$ 6 (aperture). It is a curious fact that aperture photometry on 
only one frame per night leads to relative fluxes in rough agreement with a constant level. However, more 
refined measurements (aperture or PSF fitting on several frames per night) reveal the variability of one 
("s1" or "s2") or both stars. 

\begin{figure}
\centering
\includegraphics[angle=-90,width=8cm]{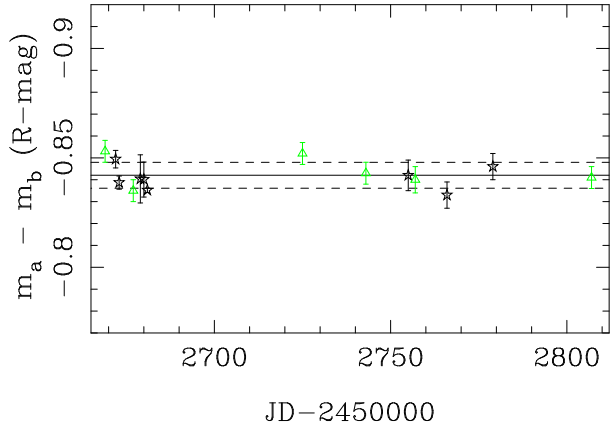}
\caption{Maidanak (open star symbols) and Wise (open triangles) light curves $y_a$ in the $R$ 
filter (winter--spring of 2003). The solid line represents the global mean value and the two dashed lines 
describe the rms fluctuation of the measurements. This rms variation agrees with the typical uncertainty, 
so "a" and "b" seem to be non--variable sources.}
\end{figure}

\begin{figure}
\centering
\includegraphics[angle=0,width=10cm]{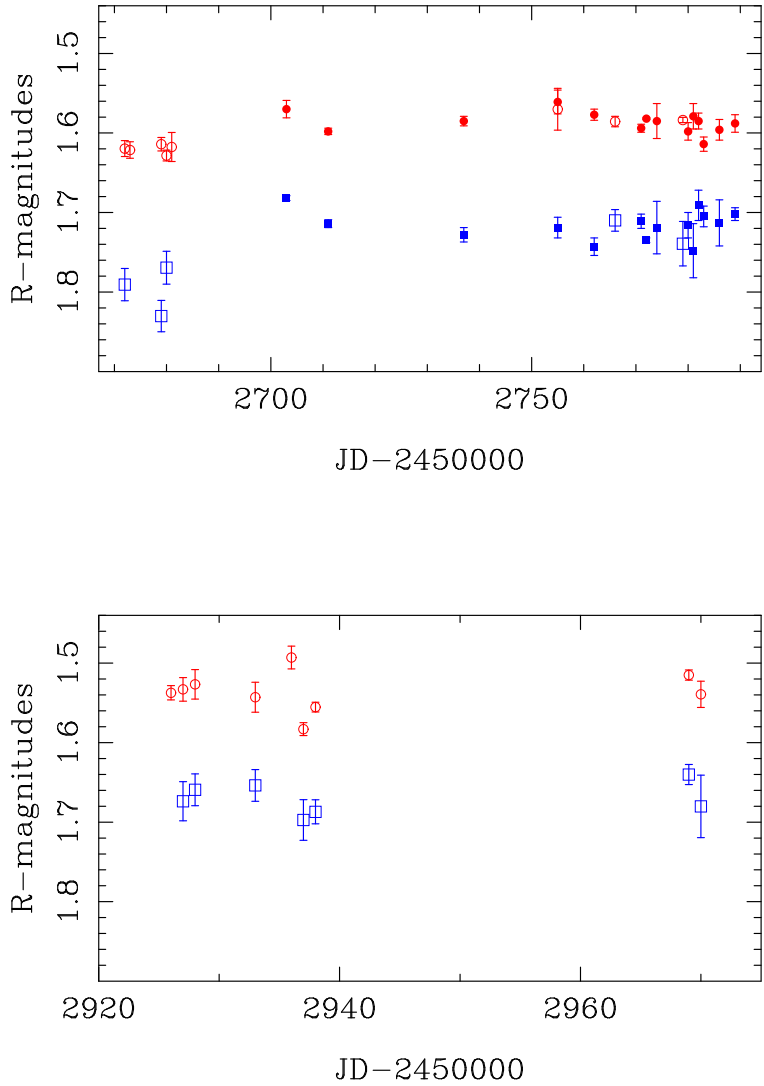}
\caption{Global $R$--band fluxes of SBS 0909+532 in 2003. The open circles (Maidanak) and filled circles 
(Calar Alto) are the fluxes $y_A$, whereas the open squares (Maidanak) and filled squares (Calar Alto) are 
the relative fluxes $y_B -$ 0.45 mag. The top panel contains the results in the winter--spring of 2003 and 
the bottom panel includes the results in the autumn of 2003.}
\end{figure}

\begin{figure}
\centering
\includegraphics[angle=0,width=10cm]{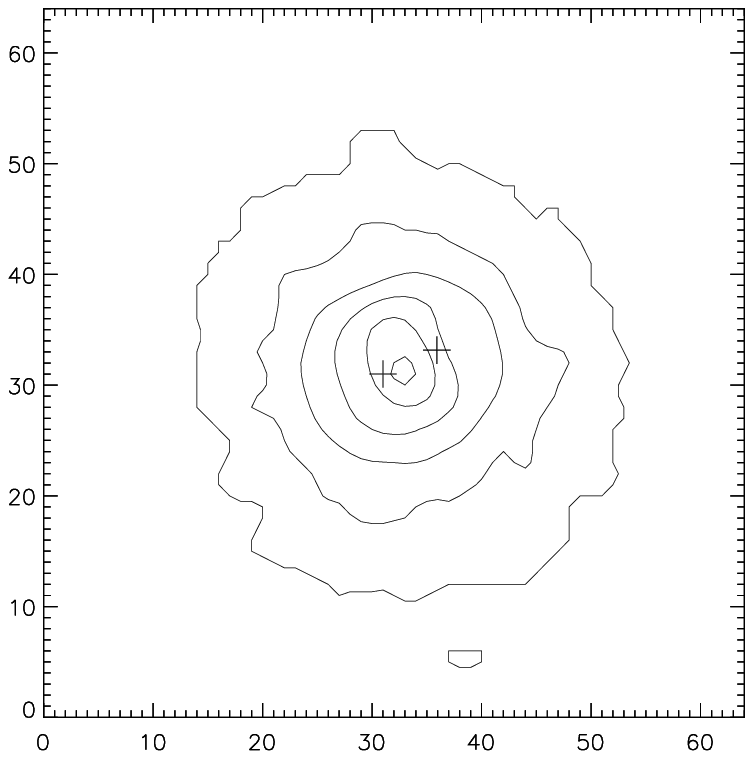}
\caption{Galaxy reconstruction obtained from selected Maidanak frames. Although the double quasar has been
subtracted here, the positions of the components are labeled with two crosses: A is on the left and B is on 
the right, and the separation between both crosses is of 1\farcs1$-$1\farcs2. The galaxy contours cover a 
region of $10\arcsec \times 10\arcsec$.}
\end{figure}

\subsection{Maidanak frames and global $R$--band light curves of SBS 0909+532}

In the case of the $R$--band Maidanak observations, in order to derive the relative fluxes of the 
components of SBS 0909+532, we also use a direct PSF fitting. For a given frame, after to obtain a first 
estimate of the free parameters (initial solution), the fit is refined through an iterative procedure, 
which works as the CLEAN algorithm ({\O}stensen 1994). The iterative task is done with each individual 
image, and the solutions converge after a few cycles. For each night, we take all the available images 
and obtain the mean values of $y_A$ and $y_B$. From the standard deviation of the means, we also derive 
the errors in $y_A$ and $y_B$. In agreement with the criteria in subsection 3.1, only fluxes with errors 
less than or equal to 40 mmag are considered. Apart from the analysis of the lens system, using aperture 
photometry, we also measure $y_a$. The relative fluxes $y_a$ are depicted in Figure 7 (open star symbols). 
The Maidanak measurements in the first semester of 2003 and the six Wise data of $y_a$ (open triangles; 
see here above) are tightly distributed around $-$ 0.842 mag (solid line in Fig. 7). The rms fluctuation 
of the data is only of $\sim$ 6 mmag (see the dashed lines in Fig. 7), which is consistent with the 
typical error of the measurements. The $y_a$ results in Fig. 7 suggest that both "a" and "b" are 
non--variable objects. Through 2003 (first and second semesters) we do not find any evidence in favour 
of variability of the "a--b" stars. 

We show our global $R$--band light curves of SBS 0909+532 in Figure 8. The open circles (Maidanak) and 
filled circles (Calar Alto) are the measurements of $y_A$, whereas the open squares (Maidanak) and filled 
squares (Calar Alto) are the values of $y_b -$ 0.45 mag. We have 31 points for the A component (circles) 
and 26 points for the B one (squares). The top panel of Fig. 8 contains the results in the winter--spring 
of 2003 and the bottom panel of Fig. 8 includes the results in the autumn of 2003. For each component we 
test the existence of a bias between the Calar Alto and Maidanak fluxes, e.g., $\beta_A = y_A$ (Calar 
Alto) $- y_A$ (Maidanak). Very small biases of $\beta_A$ = + 15 mmag and $\beta_B$ = $-$ 30 mmag are 
found, and these corrections are taken into account to make the global records in Fig. 8. The biases are
derived from the comparison between the Maidanak fluxes in a thirty day period (from day 2750 to day 
2780) and the Calar Alto fluxes at equal or close dates (see the top panel of Fig. 8). 

To roughly estimate the contaminations from the direct PSF fitting technique, we take some of our best 
Maidanak images (in terms of seeing conditions, $FWHM \sim$ 1 $\arcsec$) in the $R$ band. A zoom--in of
one of these best frames is shown in Fig. 2. Firstly, we combine the selected frames and derive a 
numerical model of the galaxy from a regularizing algorithm. To produce a more stable reconstruction, 
the real galaxy profile is assumed to be close to the Sersic profile (Koptelova et al. 2005). Our 
deconvolution method differs only slightly from the former deconvolution techniques by Magain, Courbin 
\& Sohy (1998) and Burud et al. (1998). Figure 9 presents the galaxy reconstruction obtained from the 
stack of the $R$--band selected frames. The box in Fig. 9 is 16\farcs6 on a side. The positions of the 
components are labeled with two crosses: A is on the left and B is on the right. The innermost contours 
are circular--elliptical rings, whereas the outermost contours show a less definite shape. Secondly, 
the selected frames are fitted to a photometric model that includes the galaxy brightness. Therefore, 
we are able to infer clean relative fluxes of A and B (without contamination by galaxy light) and to 
compare them with the contaminated ones (from direct PSF fitting). As result of the comparison, we 
report typical (averaged) contaminations of 18.8 mmag and 4 mmag for the A and B components, 
respectively. These very weak contaminations are in reasonable agreement with our preliminary 
considerations in the beginning of this section, and are taken into account in the measurement of the 
$R$--band flux ratio in Section 5. 

\section{Time delay}

To calculate the time delay between both components of SBS 0909+532, we use the $R$--band 
brightness records corresponding to the winter--spring of the year 2003. The $R$--band 
records are more densely populated than the $V$-band ones. Moreover, the $R$--band time 
coverage in the winter--spring of 2003 (about 120 days) is longer than the time coverage in
the autumn of 2003 (about 50 days). Thus we focus on the $R$--band data from day 2670 to day 
2790, i.e., 22 points in the A component and 19 points in the B component (see the top panel 
of Figure 8). There are different number of points for component A and component B because we only 
consider fluxes with uncertainties below 40 mmag (see Section 3). As the B component is fainter, 
its photometric uncertainties are larger and the number of final data is smaller. The new light 
curves are characterized by a mean sampling rate of one point each six days.

Once we have the data set, a suitable cross--correlation technique is required. Here we mainly
use the $\chi^2$ minimization (e.g., Kundi\'c et al. 1997) and the minimum dispersion ($D^2$) method 
(Pelt et al. 1994, 1996). However, although other techniques are probably less robust than the 
$\chi^2$ and $D^2$ ones (doing a first delay measurement, without a previous empirical determination), 
we also tentatively explore the modified cross--correlation function (MCCF) technique (Beskin \& 
Oknyanskij 1995; Oknyanskij 1997). The MCCF combines properties of both standard cross--correlation
functions: the CCF by Gaskell \& Spark (1986) and the DCF by Edelson \& Krolik (1988). We begin our 
analysis using the $\chi^2$ method, which is based on a comparison between the light curve $y_A$ (or 
$y_B$) and the time shifted light curve $y_B$ (or $y_A$). For a given lag, one can find the magnitude 
offset that minimizes the $\chi^2$ difference. From a set of lags, it can be derived a set of minima 
(of $\chi^2$), which permits to make a $\chi^2$ spectrum: $\chi^2$ vs lag. The best solution of the 
delay is the lag corresponding to the minimum of the $\chi^2$ spectrum. In general, the shifted epochs 
$t'_B$ (or $t'_A$) do not coincide with the unchanged epochs $t_A$ (or $t_B$), so we estimate the 
values of $y_A(t'_B)$ (or $y_B(t'_A)$) by averaging the A (or B) fluxes within bins centered on times 
$t'_B$ (or $t'_A$) with a semiwidth $\alpha$. To average in each bin, it is appropriate the use of 
weights depending on the separation between the central time $t'_B$ (or $t'_A$) and the dates 
$t_A$ (or $t_B$) in the bin. In principle, we concentrate in the interval [$-$ 90, + 90] days, 
which includes the predicted negative delays (see Introduction) as well as a wide range of 
unlikely positive delays (positive delays are inconsistent with basic observations 
of the system).

\begin{figure}
\centering
\includegraphics[angle=-90,width=8cm]{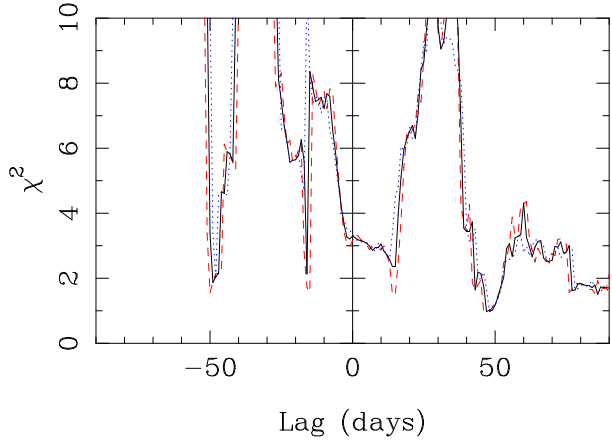}
\caption{$\chi^2$ spectra for $\alpha$ = 7 days (dashed line), $\alpha$ = 8 days (solid line) and 
$\alpha$ = 9 days (dotted line). The parameter $\alpha$ is the semiwidth of the bins in the A
component, so the three spectra represent different results of a cross--correlation with reasonable 
time--resolution (the mean sampling time is of 6 days/point).}
\end{figure}

\begin{figure}
\centering
\includegraphics[angle=-90,width=8cm]{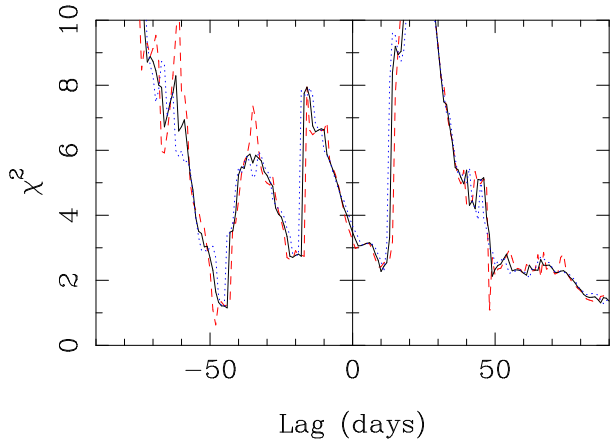}
\caption{$\chi^2$ spectra for $\alpha$ = 9 days (dashed line), $\alpha$ = 10 days (solid line) and 
$\alpha$ = 11 days (dotted line), where $\alpha$ is the semiwidth of the bins in the B component.}
\end{figure}

\begin{figure}
\centering
\includegraphics[angle=0,width=10cm]{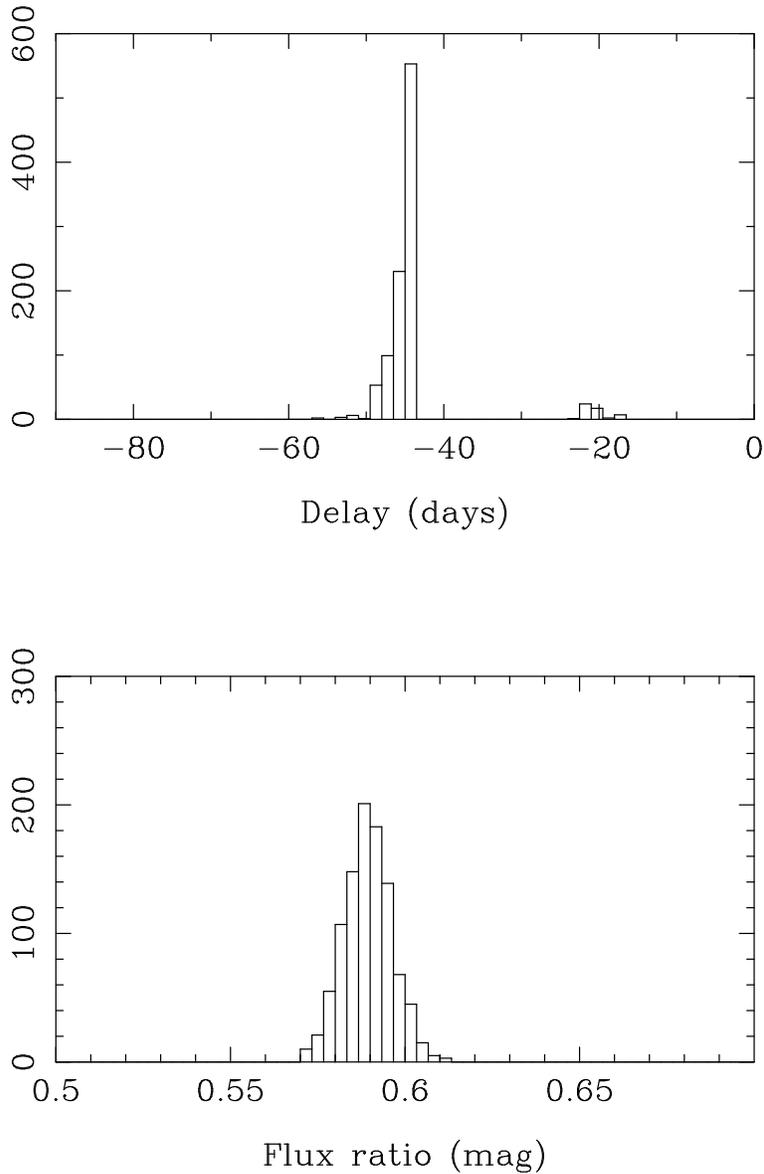}
\caption{Histograms from 1000 repetitions and the $\chi^2$ minimization (bins in B and $\alpha$ = 
10 days). {\it Top panel}: best solutions of the time delay. {\it Bottom panel}: best solutions of the 
magnitude offset (flux ratio). These distributions are consistent with a delay of about one and a half 
months and a time--delay--corrected flux ratio of about 0.59 mag.}
\end{figure}

\begin{figure}
\centering
\includegraphics[angle=0,width=10cm]{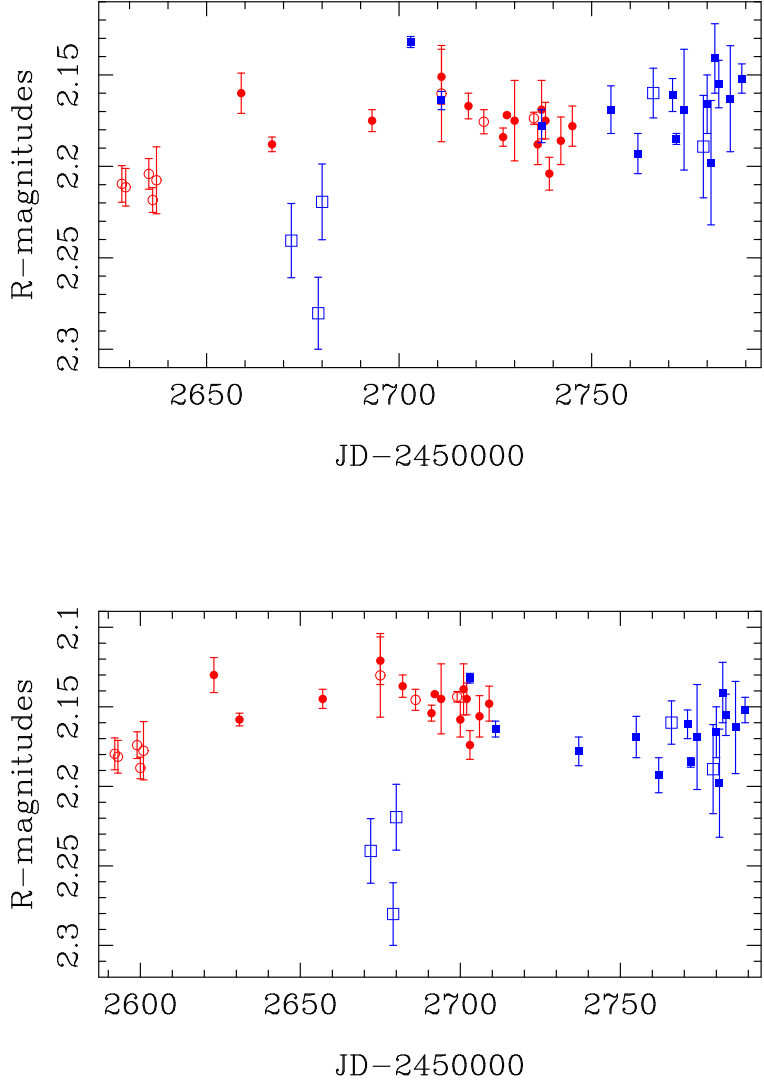}
\caption{Comparison between the shifted A light curve (circles) and the B light curve (squares).
We use a $\chi^2$ minimization, bins in B and $\alpha$ = 10 days. {\it Top panel}: best solution 
($\Delta \tau_{BA}$ = $-$ 44 days, $\Delta m_{BA}$ = 0.59 mag). {\it Bottom panel}: solution for $\Delta 
\tau_{BA}$ = $-$ 80 days ($\Delta m_{BA}$ = 0.56 mag).} 
\end{figure}

\begin{figure}
\centering
\includegraphics[angle=-90,width=8cm]{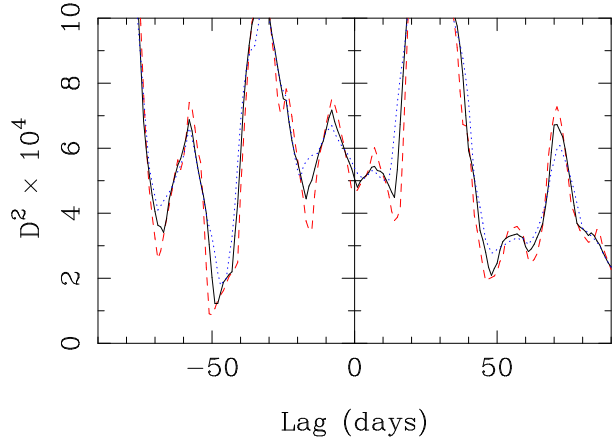}
\caption{Dispersion spectra for $\delta$ = 7 days (dashed line), $\delta$ = 9 days (solid line) and 
$\delta$ = 11 days (dotted line). The parameter $\delta$ is the decorrelation length associated with the 
$D^2_{4,2}$ spectra (Pelt et al. 1996).}
\end{figure}

\begin{figure}
\centering
\includegraphics[angle=0,width=10cm]{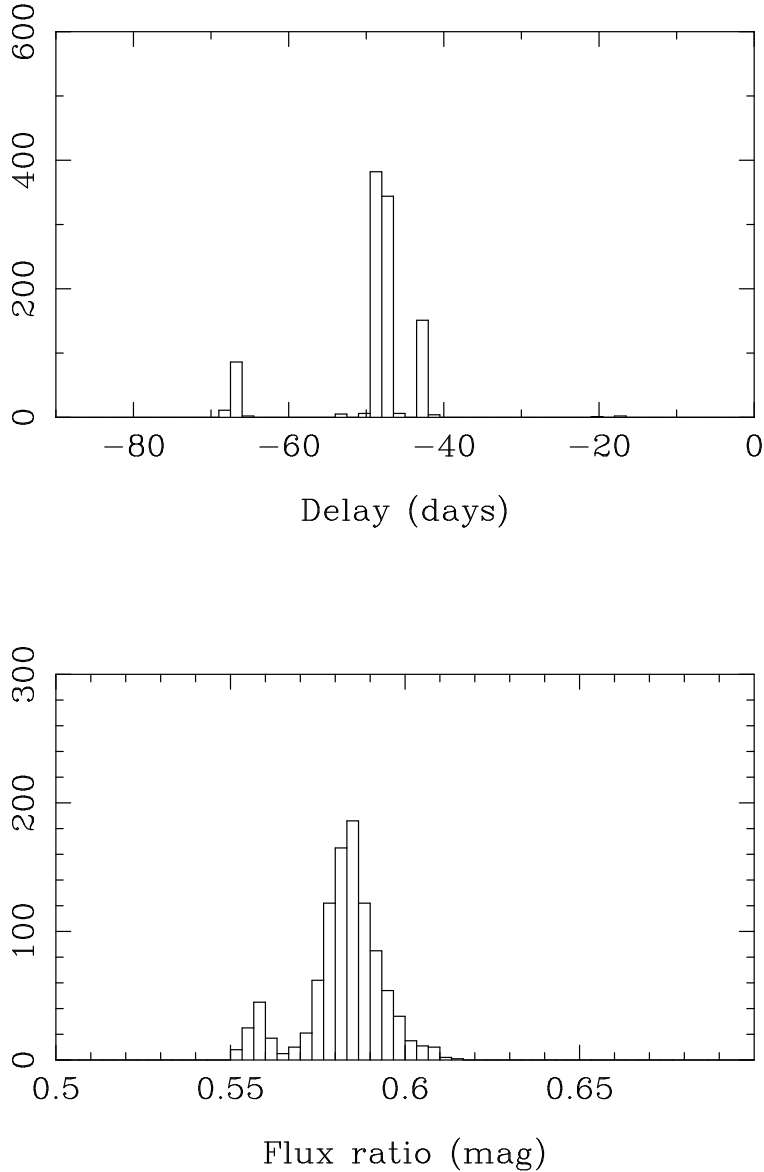}
\caption{Histograms from 1000 repetitions and the $D^2$ minimization ($\delta$ = 9 days). 
{\it Top panel}: best solutions of the time delay. {\it Bottom panel}: best solutions of the 
magnitude offset (flux ratio). These distributions are again consistent with an 1.5--month delay and
and a time--delay--corrected flux ratio of about 0.58--0.59 mag.}
\end{figure}

Firstly, the curve $y_A$ and the time shifted curve $y_B$ are compared with each other 
(using bins in the A component). In order to work with a reasonable time--resolution, we use 
$\alpha$ values less than or equal to two times the mean sampling time, i.e., $\alpha \leq$ 
12 days. The $\chi^2$ value roughly grows with the size of the bin, and $\chi^2 \sim$ 1 for 
$\alpha$ = 7$-$9 days. For $\alpha$ = 7$-$9 days, there are best solutions $\Delta \tau_{BA}$ 
= $+$ 46$-$48 days ($\chi^2$ = 0.97$-$0.98), and we show the corresponding spectra in Figure 
10. We have drawn together the spectra for $\alpha$ = 7 days (dashed line), $\alpha$ = 8 days 
(solid line) and $\alpha$ = 9 days (dotted line). Apart from the main minima close to + 50 days, 
there are other secondary minima at negative and positive lags. In Fig. 10, two secondary
minima seem to stay significant for all the bin sizes: the minima close to $-$ 50 days and the
probable edge effects at + 80$-$90 days. We also compare the curve $y_B$ and the time shifted 
curve $y_A$, using bins in the B component. For $\alpha$ = 10 days, we obtain a best solution
$\Delta \tau_{BA}$ = $-$ 44 days ($\chi^2$ = 1.15). Smaller and larger bins lead to solutions 
characterized by $\chi^2 <$ 0.7 and $\chi^2 \geq$ 1.2, respectively. In Figure 11, the solid 
line represents the spectrum for $\alpha$ = 10 days, while the dashed line represents the
spectrum for $\alpha$ = 9 days and the dotted line traces the spectrum for $\alpha$ = 11 days. 
Main minima in the interval $-$ 40$-$50 days appear in all these cases. Unfortunately, 
important signals at positive lags and probable border effects at + 80$-$90 days are again
included in the complex spectra. The important structures at positive lags in Figs. 10$-$11 are 
probably caused by artifacts in the cross--correlation, so they have no physical origin, but 
are due to the 10/20--day gaps and the moderate variability of the components. Therefore, 
taking $\alpha$ = 10 days (bins in the B component) and a negative range [$-$ 90, 0] days, we 
try to determine a pre--conditioned time delay. 

In order to derive uncertainties, we follow a simple approach. We make one repetition 
of the experiment by adding a random quantity to each original flux in the light curves. The 
random quantities are realizations of normal distributions around zero, with standard 
deviations equal to the errors of the fluxes. We can make a large number of repetitions, 
and thus, obtain a large number of $\Delta \tau_{BA}$ values. The true value will be included 
in the whole distribution of measured delays. From the $\chi^2$ minimization (bins in B and 
$\alpha$ = 10 days) and 1000 repetitions, we obtain the histograms in Figure 12. Regarding the 
distributions in the top panel (delays) and bottom panel (flux ratios) of Fig. 12, the main 
features lead to measurements $\Delta \tau_{BA}$ = $-$ 45 $^{+ 1}_{-11}$ days and $\Delta 
m_{BA}$ = 0.590 $\pm$ 0.014 mag (95\% confidence intervals). We note that the main delay peak 
is asymmetric, so 55\% of the repetitions correspond to $-$ 44$-$45 days, whereas 40\% of the 
repetitions correspond to values $<$ $-$ 45 days. The secondary delay peak (around $-$ 20 days) 
represents about 5\% of the repetitions and is associated with the secondary minima in the 
negative region of Fig. 11. Therefore, the distribution in the top panel of Fig. 12 permits a 
95\% estimation of the time delay of SBS 0909+532. 

In Figure 13 (top panel), the A light curve (circles) shifted by the optimal values of the 
time delay and the magnitude offset (time--delay--corrected flux ratio), and the unchanged 
B light curve (squares) are plotted. The cross--correlation using bins in the B component
($\alpha$ = 10 days) indicates that the initial variations in the brightness of B reasonably 
agree with the final fluctuations in the brightness of A. The overlap for a delay of $-$ 80 
days (e.g., Saha et al. 2005) also appears in the bottom panel of Fig. 13. However, this last 
time delay is clearly rejected by the observations, since the $\chi^2$ value is larger than 
10 ($\chi^2 \sim$ 18). 

To confirm the results from the $\chi^2$ minimization, we also use the dispersion spectra 
introduced by Pelt et al. (1994, 1996). The basic idea is a combination of $y_A$ and $y_B$ 
into one global record for every lag $\tau$ and magnitude offset $m_0$ by taking all the 
values of $y_A$ as they are and shifting the values of $y_B - m_0$ by $\tau$. For each $\tau$  
one can find the $m_0$ value that minimizes a dispersion estimate $D^2(\tau,m_0)$, so a 
dispersion spectrum $D^2(\tau)$ can be made in a direct way. We focus on the $D^2_{4,2}$ 
spectra that are called $D^2$ for simplicity (see Pelt et al. 1996 for details). This technique 
incorporates a decorrelation length ($\delta$), where $\delta$ plays a role similar to that 
of $\alpha$ in the $\chi^2$ method. Considering reasonable values of $\delta$ (from 7 to 11 
days, see here above), we are able to make some interesting spectra. In Figure 14 we have 
plotted together the spectra for $\delta$ = 7 days (dashed line), $\delta$ = 9 days 
(solid line) and $\delta$ = 11 days (dotted line). Although there are main minima in the interval 
$-$ 40$-$50 days, there are also significant signals at positive lags and probable border 
effects at + 90 days. In the negative region of Fig. 14, a secondary minimum around $-$ 70 days
appears. Using $\delta$ = 9 days and a negative range [$-$ 90, 0] days, we carry out a second
pre--conditioned measurement of the time delay. The uncertainties are deduced from 1000  
repetitions of the experiment (see here above), and the relevant histograms are shown in 
Figure 15. While the top panel contains the distribution of delays, the bottom panel traces the
distribution of flux ratios. Through the distributions in Fig. 15, we obtain that $\Delta 
\tau_{BA}$ = $-$ 48 $^{+ 7}_{-6}$ days and $\Delta m_{BA}$ = 0.585 $\pm$ 0.020 mag (90\% 
confidence interval). These $D^2$ results strengthen the conclusions from the $\chi^2$ 
technique. A marginal measurement (10\% confidence interval) of $\Delta \tau_{BA}$ = $-$ 
67 $^{+ 1}_{-2}$ days and $\Delta m_{BA}$ = 0.558 $^{+ 0.007}_{-0.008}$ mag is also possible.
However, both this possibility and the $\chi^2$ result of around $-$ 20 days are probably 
related to the presence of gaps and the absence of strong variability in the light curves. 

A MCCF technique (Beskin \& Oknyanskij 1995; Oknyanskij 1997) is also explored. The MCCF is 
a modification of the standard cross--correlation functions (CCF and DCF). When this MCFF is
applied to our data in the lag interval [$-$ 60, + 60] days, the maximum correlation 
coefficient (0.907) corresponds to a lag of $-$ 45 days. This last result basically agrees 
with the $\chi^2$ and dispersion spectra in Figs. 11 and 14. 

\section{Conclusions}

Nowadays several groups are trying to coordinate the rich but scattered research potential 
in the field of gravitationally lensed quasar monitoring. The goals are to rationalize the 
astronomical work and to catalyze big scientific collaborations so that the astrophysics
community can get a significant progress in the understanding of the central engine in 
lensed quasars, the structure of the lensing galaxies and the physical properties of the 
Universe as a whole. Some examples about that are the Astrophysics Network for Galaxy 
LEnsing Studies (ANGLES, http://www.angles.eu.org/), the Cosmic Lens All-Sky Survey (CLASS, 
http://www.aoc.nrao.edu/$\sim$smyers/class.html) and the COSmological MOnitoring of 
GRAvItational Lenses (COSMOGRAIL, http://www.cosmograil.org/).
The University of Cantabria group (Spain), three groups of the former Soviet Union 
(Institute of Astronomy of Kharkov National University, Ukraine, Sternberg Astronomical 
Institute, Russia, and Ulug Beg Astronomical Institute of Uzbek Academy of Science, 
Uzbekistan) and the Tel--Aviv University group (Israel) are also carrying out a series of 
initiatives to better exploit the recent individual monitoring campaigns as well as to 
solidify some future common project. In this paper we present the first collaborative 
programme on the variability of the double quasar SBS 0909+532A,B. The $VR$ observations of 
the system and the field stars were made with three modern ground--based telescopes in the 
year 2003. 

The SBS 0909+532c star (N23210036195 in the GSC2.2 Catalogue) at ($\alpha$, $\delta$) = 
(09:12:53.59, +52:59:39.82) in J2000 coordinates is found to be variable, with two 
different levels of flux. The $VR$ gap between the low--state and the high--state is of
80--100 mmag, and the low--state lasts about one month. In the high--state the star also
seems to vary, but these small--amplitude variations are not so significant as the gap 
between states. We want to remark the variability of this nearby star ("c" star), and 
to encourage colleagues to follow-up its fluctuations and identify the kind of variable 
source. The "c" star cannot be used as the reference object (differential photometry), 
because it introduces a zero--lag global correlation between the light curves of the
quasar components A and B. However, the "a--b" nearby stars are non--variable sources, 
and we choose the "b" star as the reference candle. On the other hand, the "s1" and "s2" 
stars are relatively far objects, which were proposed as good references in a previous 
analysis (Nakos et al. 2003). However, the new $R$--band light curve $m_{s2} - m_{s1}$ 
reveals the variability of one ("s1" or "s2") or both stars. This variability could be 
either a very rare phenomenon or a consequence of doing more refined measurements 
(aperture or PSF fitting on several frames per night). We warn about the possible 
problems with this pair of stars and think it merits more attention. The point--spread 
function (PSF) fitting methods permit to resolve the two components of the quasar and 
to derive the $VR$ light curves of each component. These new $VR$ light curves represent 
the first resolved brightness records of SBS 0909+532. Although the $V$--band curves are 
interesting, the $R$--band records seem more reliable and are more densely populated. 
The $R$--band curves show a moderate variability through 2003, and the observed fluctuations 
are promising for different kinds of future studies. 

To estimate the time delay between the components of SBS 0909+532, we use an 120--day piece
of the $R$–-band brightness records, and $\chi^2$ and dispersion ($D^2$) techniques. The 
cross--correlation of the two light curves (A and B) leads to complex $\chi^2$ spectra. 
However, assuming that the quasar emission is observed first in B and afterwards in A, or in
other words, $\Delta \tau_{BA} <$ 0 (in agreement with basic observations of the
system), 95\% measurements $\Delta \tau_{BA}$ = $-$ 45 $^{+ 1}_{-11}$ days and 
$\Delta m_{BA}$ = 0.590 $\pm$ 0.014 mag are inferred from 1000 repetitions of the 
experiment (synthetic light curves based on the observed records). From the $D^2$ 
minimization (Pelt et al. 1996) and 1000 repetitions, we also obtain 90\% measurements $\Delta 
\tau_{BA}$ = $-$ 48 $^{+ 7}_{-6}$ days and $\Delta m_{BA}$ = 0.585 $\pm$ 0.020 mag. The $D^2$ 
uncertainties are derived under the already mentioned assumption that $\Delta 
\tau_{BA}$ is negative. There is a clear agreement between the results from both techniques, so
a delay value of about one and a half months is strongly favoured. Our light curves rule 
out a delay close to three months, which has been claimed in a recent analysis (Saha et 
al. 2005). When we measure the time delay of the system, we simultaneously derive 
the time--delay--corrected flux ratio (at the same emission time) in the $R$ band. This 
quantity, $\Delta m_{BA}$ = $m_B(t + \Delta \tau_{BA}) - m_A(t)$, is contaminated by light of 
the lens galaxy, and taking into account the weak contaminations of A and B (see the end of 
subsection 3.2), the totally corrected $R$--band flux ratio is 0.575 $\pm$ 0.014 mag. We 
remark that our final $R$ flux ratio is in total agreement with the rough (uncorrected by the 
time delay and the contamination by galaxy light) measurement by Kochanek et al. (1997): 0.58 
$\pm$ 0.01 mag. To properly determine a flux ratio, one must use clean fluxes at the same 
emission time, i.e., fluxes at different observation times and without contamination 
(Goicoechea, Gil--Merino \& Ull\'an 2005). Only for particular cases (e.g., faint lens galaxy, 
short delay and moderate variability), it may be reasonable to use direct fluxes.  

In order to get a reasonably good value of $\chi^2$, we do not need to introduce a time 
dependent magnitude offset or a complex iterative procedure (e.g., Burud et al. 2000; Hjorth 
et al. 2002), i.e., only a delay and a constant offset are fitted. This is a strong point of 
the analysis. The agreement between the results from different techniques is another strong 
point. However, the new measurements have some weak points that we want to comment here. The 
weakest point is the relatively poor overlap between the A and B records, when the A light 
curve is shifted by the best solutions of the time delay and the magnitude offset (e.g., see 
the top panel of Fig. 13). Moreover, we carry out pre--conditioned measurements, since a 
negative interval [$-$ 90, 0] days is considered in the estimation of uncertainties (component 
B leading component A). This second weak point is related to the presence of 10/20--day gaps 
and the moderate variability of the components, which does not permit to fairly rule out 
positive delays. We nevertheless remark that the negative interval is in good agreement with 
the predictions by Leh\'ar et al. (2000) and Saha et al. (2005), and we find $\chi^2$ and $D^2$
minima around $-$ 45 days when the observed data and both negative and positive lags are taken 
into account (see Figs. 11 and 14). Of course, as any another first determination of a time delay, 
the 1.5--month value should be confirmed from future studies.

Forty years ago, Refsdal (1964) suggested the possibility of determining the current 
expansion rate of the Universe (Hubble constant) and the masses of the galaxies from the 
time delays associated with extragalactic gravitational mirages. More recently, for a 
singular isothermal ellipsoid (SIE), Koopmans, de Bruyn \& Jackson (1998) found that 
the time delay can be cast in a very simple form, depending on basic cosmological 
parameters, redshifts and image positions. The relevant image positions are the 
positions with respect to the centre of the main lens galaxy, and the SIE delay is 
similar to the delay for a singular isothermal sphere (SIS). In principle, a singular 
density distribution is justified because a small core radius changes the time delay 
negligibly, and only a small core radius seems to be consistent with the absence of 
a faint central image (e.g., Kochanek 1996). Moreover, individual lenses and lens 
statistics are usually consistent with isothermal models (e.g., Witt, Mao \& Keeton 
2000 and references therein), so it is common to adopt an isothermal profile. Witt, 
Mao \& Keeton (2000) showed that an external shear changes the simple SIS time delay 
in proportion to the shear strength. For two--image lenses that have a small shear and 
images at different distances from the centre of the lens, the shear should have a 
small effect on the time delay. Thus, when one has accurate measurements of image 
positions, redshifts and time delay, it is viable an accurate estimation of $H_0$ 
(using complementary information on the matter/energy content of the Universe).

Very recently, Kochanek (2002) also presented a new elegant approach to the subject. 
He modelled the surface density locally as a circular power law, with a mean surface 
density $<\kappa>$ in the annulus between the images. Expanding the time delay as a 
series in the ratio of the thickness of the annulus to its average radius, it is 
derived a delay that is proportional to the SIS time delay. The zero--order expansion 
term consists of the SIS delay and a multiplicative factor $2(1 - <\kappa>)$. Kochanek 
also incorporated the quadrupoles of an internal shear (ellipsoid) and an external 
shear. However, for two-image lenses where the images lie on opposite sides of the 
lens, the delay depends little on the quadrupoles. This novel perspective is useful 
to infer $<\kappa>$ from observations of the lens system (time delay, image positions 
and redshifts) and complementary cosmological data (expansion and matter/energy 
content of the Universe).

For SBS 0909+532, although the redshifts are very accurately known and the time delay 
is now tightly constrained (or at least there is a first accurate estimation to be 
independently confirmed), the inaccurate position of the main lens galaxy does not 
permit an accurate measurement the cosmic expansion rate and the surface density of the 
main deflector. We have $H_0 \propto \theta_B^2 - \theta_A^2$ and $1 - <\kappa> 
\propto (\theta_B^2 - \theta_A^2)^{-1}$, where $\theta_A$ and $\theta_B$ are the image
angular positions with respect to the centre of the main lens galaxy. On the other
hand, using the astrometry in Table 3 of Leh\'ar et al. (2000), it is easy to obtain 
$\theta_B^2 - \theta_A^2$ = 0.4 $\pm$ 0.2. Thus we conclude that the accuracy in 
$\theta_B^2 - \theta_A^2$ is only 50\%, indicating the necessity of new accurate 
astrometry of SBS 0909+532.

\begin{acknowledgements}
The UC members are indebted to J. Alcolea (Observatorio Astron\'omico Nacional, Spain) for 
generously granting permission to operate the 1.52 m Spanish telescope at Calar Alto 
Observatory (EOCA) in March--June 2003. This 4--month season was supported by Universidad de 
Cantabria funds and the Spanish Department for Science and Technology grant AYA2001-1647-C02. 
AU thanks the Departamento de F\'isica Te\'orica y del Cosmos de la Universidad de Granada (E. 
Battaner) for hospitality during the observational season. The post--observational work and the 
spreading of results are supported by the Department of Education and Science grants 
AYA2002-11324-E, AYA2004-20437-E and AYA2004-08243-C03-02. We acknowledge the use of data 
obtained by the SAI group headed by B. Artamonov. We are also indebted to D. Maoz and E. Ofek 
for providing us with the Wise frames of SBS 0909+532. APZ is grateful for the support of 
the Science and Technology Center of Ukraine (STCU), grant U127k. The observational work by the 
UBAI group (TA and OB) at Mt. Maidanak was supported by the German Research Foundation (DFG), 
grant 436 UZB 113/5/0-1. We are also grateful to the referee for several helpful comments.
We acknowledge support by the European Community's Sixth Framework Marie Curie Research Training 
Network Programme, Contract No.MRTN-CT-2004-505183 "ANGLES". The GSC-II is a joint project of the 
Space Telescope Science Institute (STScI) and the Osservatorio Astronomico di Torino (OAT). STScI 
is operated by the Association of Universities for Research in Astronomy, for the NASA under 
contract NAS5-26555. The participation of the OAT is supported by the Italian Council for 
Research in Astronomy. Additional support is provided by ESO, Space Telescope European 
Coordinating Facility, the International GEMINI project and the ESA.
\end{acknowledgements}

{}

\end{document}